\documentclass{iau}
\usepackage{graphicx}

\title[IAUS291.Intermittent PSR J1839+15] %% short title %%
{Discovery of an Intermittent Pulsar:\\ PSR J1839+15} %% full title %%

\author[Surnis \etal]  %% short author list %%
{M. P. Surnis$^1$, B. C. Joshi$^1$, M. A. McLaughlin$^2$, \and V. Gajjar$^1$}

\affiliation{$^1$National Centre for Radio Astrophysics, Pune, India. \\ email: {\tt mpsurnis@ncra.tifr.res.in} \\[\affilskip]
$^2$Department of Physics, West Virginia University, Morgantown, USA.}

\pubyear{2012}
\volume{291}  %% insert here IAU Symposium No.
\jname{\mbox{Neutron Stars and Pulsars: Challenges and Opportunities after 80 years}}
\editors{J. van Leeuwen, ed.} 
\begin{document}

\maketitle

%% -- Abstract ----------------------------------
\begin{abstract}
We report the discovery of a new pulsar PSR J1839+15, having a period of 549 ms and a DM of 68 pc-cm$^{-3}$. We also present its timing solution and report the intermittent behaviour of its radio emission.

\end{abstract}

% add below any authors, subjects and objects for indexing 
%   add more lines if necessary
%   but leave all lines commented out
%\index[author]{LastName1, Initials|textbf}
%\index[author]{LastName2, Initials|textbf}
%\index[subject]{Keyword1}
%\index[subject]{Keyword2}
%\index[object]{Object1}
%\index[object]{Object2}

\firstsection % if your document starts with a section,
\section{Introduction}
A blind search for pulsars along the Galactic plane covering 10\% of the region between Galactic longitude 45$^{\circ} <$ l $<$ 135$^{\circ}$ and Galactic latitude 0$^{\circ} < |$b$| <$ 5$^{\circ}$ was recently carried out with the Giant Metrewave Radio Telescope (GMRT). It was named \lq GMRT Galactic Plane Pulsar and Transient Survey\rq. The survey was carried out at 325 MHz with a bandwidth of 16 MHz, divided into 256 filterbank channels. Each field (circular region of radius $\sim$ 1$^{\circ}$) was observed for 1800 s with a sampling time of 256 $\mu$s.\\

\paragraph{}
The advantage of observing at 325 MHz was the wide field of view ($\sim$ 4 deg$^2$) paired with a sensitivity of 0.6 mJy. The observations were carried out using the incoherent array (IA) mode of the GMRT. The data were written to magnetic tapes. They were extracted to network attached storage (NAS) disks of the high performance computing cluster having 64 dual core nodes. Pulsar search was done using SIGPROC\footnote{\tt www.sigproc.sourceforge.net} with extensive RFI mitigation algorithms written by one of the authors. The trial DM range used for the search was 0$-$1200 pc-cm$^{-3}$. The analysis was parallelised such that the DM search for a particular field was divided among the nodes. The results were written back to the NAS disks from where, they could be copied to other machines. The candidate plots thus generated, were manually scrutinised for identifying good candidates.\\

\paragraph{}
The follow-up timing observations continued with a new software back-end at GMRT. This provided 512 filterbank channels across a bandwidth of 33 MHz with 122.88 $\mu$s sampling. The integration time was 1800 s. The timing analysis was done using TEMPO2\footnote{\tt www.atnf.csiro.au/research/pulsar/tempo2} (\cite{H06}).

\section{Results}
\begin{figure}[h]
\begin{center}

\includegraphics[scale=0.55]{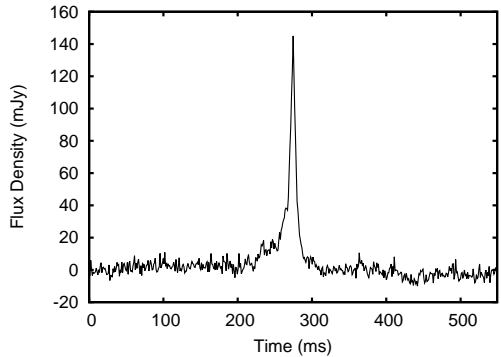}
\caption{The accumulated pulse profile for PSR J1839+15 at 325 MHz. The ordinate was scaled using radiometer equation.}
\label{accprof}

\end{center}
\end{figure}

PSR J1839+15 came out as a strong candidate and was successfully confirmed in later follow-up observations. The accumulated profile from 23 detections on the pulsar is shown in Figure \ref{accprof}. It shows a very narrow peak. A weaker second component can be seen just before the main pulse. The region before this component may also have another, still weaker component buried in the noise. Overall, it may consist of two or more components although polarization studies are required to confirm the same. The timing solution obtained so far is given in Table \ref{param}\\

\begin{table}[h!]
\begin{center}
\begin{tabular}{l r}
\hline
RA                                       & 18h39m06.6(2)s\\
DEC                                      & +15$^{\circ}$06'57(6)"\\
P                                        & 0.54916053388(8) s\\
\.{P}                                    & 2.613(6) $\times$ 10$^{-14}$ s/s\\
DM                                       & 68.1(8) pc-cm$^{-3}$\\
Characteristic age $\tau$ = P/2\.{P}     & 0.33 Myr\\
Surface Magnetic Field B$_{S}$           & 3.83 $\times$ 10$^{12}$ G\\
DM Distance = DM/n$_{e}$                 & 3 kpc$^a$\\ 
\hline
$^a$DM distance calculated using the model given by \cite{C02} 
\end{tabular}     
\caption{Important parameters of PSR J1839+15}
Numbers in brackets indicate 2$\sigma$ errors as reported by TEMPO2 in the last digit of the given value. The error on the DM comes from local search done on the time series data.
\label{param} 
\end{center}
\end{table}  

\begin{figure}[t]
\begin{center}

\includegraphics[scale=0.55]{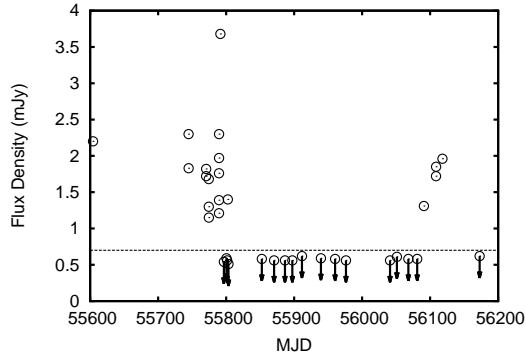}
\caption{Estimated mean flux density of PSR J1839+15 as a function of MJD. Downward arrows indicate 8$\sigma$ flux density limits on non-detections. The dashed horizontal line indicates 98\% confidence limit for the lowest expected flux density given the variations in the observed flux densities.}
\label{fluvar}

\end{center}
\end{figure}

\paragraph{}
During the follow-up timing observations, the pulsar could not be detected for 278 days from 30$^{th}$ August 2012 to 13$^{th}$ June 2012. It could be detected regularly then onwards until on 2$^{nd}$ September 2012, when it was again not detected. The estimated mean flux densities for the detections are plotted in Figure \ref{fluvar}. As can be clearly seen, the 8$\sigma$ upper limits on non-detections are below the 98\% confidence limit on the expected flux density. Thus, we are fairly confident that PSR J1839+15 is an intermittent pulsar with an OFF time scale of roughly 278 days. We are currently working on calculating the \.{P} in the ON and OFF states. Two different values of \.{P} would confirm this as an intermittent pulsar. There are only three more such pulsars known currently. PSR B1931+24 shows a quasi periodic ON-OFF cycle of about 30-40 days (\cite{K06}), PSR J1841$-$0500 shows an OFF time scale of 580 days (\cite{C12}) while J1832+0029 remained OFF for 650 days and 850 days in the two sampled OFF states (\cite{L12}). The reason for this behavior is not understood. Further investigations and discoveries of new members of this class may shed some light on the underlying physics.
\clearpage

\section{Discussion}
Intermittent pulsars are a rare breed of pulsars showing very long period nulls. The cause of these nulls is a mystery. To add to the already puzzling phenomenon, it was reported (\cite{K06}, \cite{C12} and \cite{L12}) that the spin-down rate is considerably higher in the ON state than in the OFF state indicating that the particle flow forming the magnetospheric currents, is different in the ON and OFF states. Another puzzling fact is that these pulsars belong to the normal pulsar population in the P$-$\.{P} diagram. The cause of cessation of radio emission altogether may be attributed to the reduced particle flow or it may just be a failure of coherent emission. These pulsars certainly challenge the current emission models and if well studied, may provide vital inputs for coming up with more realistic emission mechanisms.\\ 

\paragraph{}
Given the ON-OFF nature of these pulsars, they provide great motivation for extending the existing blind searches and embarking on new, sensitive blind searches even in the previously searched areas of sky. Assuming a typical duration of a big survey as 3 to 4 years and given the OFF state time scales of these pulsars, a rough estimate of the intermittent pulsar population may go up considerably, thus opening up the possibility of discovering many more of such objects.

\section{Acknowledgements}
We would like to thank the staff of the GMRT who have made these observations possible. GMRT is run by the National Centre for Radio Astrophysics of the Tata Institute of Fundamental Research.

\end{document}